\documentstyle[prl,psfig,aps,epsf,graphics,twocolumn]{revtex}
\begin{document}
\draft
%
%
%
%

\title{Excitons in Mott insulators}

\author{P. Wr\'obel$^{1}$ and R. Eder$^2$}

\address{$^1$ Institute for Low Temperature and Structure
Research, P. 0. Box 1410, 50-950 Wroc{\l}aw 2, Poland}
\address{$^2$ Institut f\"ur Theoretische Physik, Universit\"at W\"urzburg,
Am Hubland, 97074 W\"urzburg, Germany}

\date{\today}
\maketitle

\begin{abstract}
Motivated by  recent Raman and resonant inelastic X-ray
scattering experiments performed for Mott insulators, which suggest
formation of excitons in these systems, we present a theory of
exciton formation in the upper Hubbard band. The analysis based on
the spin polaron approach  is performed in the framework of an
effective $t$-$J$ model for the subspace of states with one doubly
occupied site. Our results  confirm the existence of excitons and
bear qualitative resemblance to experimental data despite some
simplifications in  
our approach. They prove that the basic underlying mechanism
of exciton formation is the same as that which gives rise to  binding of
holes in
weakly doped 
antiferromagnets.
\end{abstract}


The electronic properties of Mott insulators, like high
temperature superconductors in the limit of low doping, belong to
the most intriguing questions which faces contemporary condensed
matter physics. Properties of such systems are determined by the large
onsite Coulomb interaction $U$ or the charge transfer energy
$\Delta$ of removing an electron from the $p$ orbital of oxygen
and putting it into the $d$ orbital of the transition metal. The
structure of occupied electronic states has been intensively
investigated by means of the angle-resolved photoemission
spectroscopy (ARPES) and much is already known about the
electronic structure below the insulating gap. Natural probes of
unoccupied states above the gap are Raman\cite{Chenetal1997} and
resonant inelastic X-ray scattering\cite{Hasanetal2000} (RIXS).

The Raman scattering provides information about
excitations at ${\bf k}=0$. If we ignore polaritonic effects
the cross section for the Raman scattering may be obtained
directly from the golden rule. The final formula
\cite{ChubukovFrenkel1995} contains matrix elements $\langle
f|M_R|i\rangle$ evaluated between an initial state $i$ and a final
state $f$. In the case of the Mott insulator in two dimensions (2D)
described by the Hubbard model at half-filling in the limit of
strong correlations (large $U$), which is a natural framework to 
discuss such a system, the initial state $i$ is just
the ground state of the quantum antiferromagnet (AF) in 2D, while
the final state may be identified as an excitation in the AF state.
The electronic states $|n \rangle$ above the insulating gap appear
at the intermediate stage of the Raman process, which may be seen
from the explicit formula for $\langle f|M_R|i \rangle$,
\begin{eqnarray}
\langle f|M_R|i \rangle &=&\sum_n[\frac{\langle f|{\bf
j}_{k_f}\cdot \hat{\bf e}^\ast_f |n \rangle \langle n|{\bf
j}_{-k_i}\cdot \hat{\bf e}_i |i \rangle}{\epsilon_i + \hbar
\omega_i -\epsilon_n + i \delta}  \nonumber \\ &&+ \frac{\langle
f|{\bf j}_{-k_i}\cdot \hat{\bf e}_i |n \rangle \langle n|{\bf
j}_{-k_i}\cdot \hat{\bf e}^\ast_f |i
\rangle}{\epsilon_i-\epsilon_n - \hbar \omega_f  + i \delta}].
\end{eqnarray}
$\epsilon_i$ and $\epsilon_f$ are energies of the initial and
final states, while $\hbar \omega_i$, $\hat{\bf e}_i$ and $\hbar
\omega_f$, $\hat{\bf e}_f$ represent energies and polarizations of
incident and outgoing photons. The current operator for the
Hubbard model on the square lattice is given by,
\begin{equation}
{\bf j}_{k=0}=it\sum_{i,\delta,\sigma}{\bf
\delta}c^{\dag}_{i+\delta,\sigma} c_{i,\sigma}. \label{curr}
\end{equation}
The summation is carried over lattice sites $i$, vectors
$\delta$ which connect each site with its four nearest neighbors
and spin polarizations $\sigma$. The Hubbard model in the lower
Hubbard band at half-filling in the limit of strong correlations
is equivalent to the Heisenberg model which action is restricted
to the Hilbert space spanning states that correspond to
configurations in which each site is occupied by a single spin-$1/2$
fermion. The current operator (\ref{curr}) acting on the
antiferromagnetic state, shifts a fermion to a nearest neighbor
site that is already occupied, which means that it couples lower
and upper Hubbard bands (LHB, UHB). The newly created hole and the
double occupancy (DO) of a site reside on nearest neighbor sites. We
conclude, that in order to contribute some substantial weight to
the Raman spectra, the intermediate state should be a 
 bound state of some sort, which is formed by  the hole and the DO
 that plays 
the role of a
particle. In other words, we expect formation of excitons in the
UHB. The current operator at ${\bf k}=0$ is a vector and obeys a
dipole selection rule. Since the ground state of the quantum AF is
fully symmetric, the current operator couples it only with
excitons of $p$-wave symmetry.

The RIXS process may be
interpreted in a slightly similar way, which in addition requires
explicit consideration of the orbital structure of involved
states. According to a most likely scenario for the undoped cuprates
\cite{Hilletal1998,PlatzmanIsaacs1998,Abbamonteetal1999,IdeKotani1999,Tsutsuietal1999,Hasanetal2000}
$3d$ electrons interact with the $1s$ core hole created during the dipole
transition of an $1s$ electron to the $4p$ orbital which
accompanies the absorption of an incident photon. This strong
interaction gives rise to a shift of the $3d$ electrons across the gap. In
the last step, the $4p$ electron returns  to the $1s$
orbital and an outgoing photon is emitted. The RIXS process like
the Raman scattering creates a particle-hole pair in the UHB which
carries energy and in addition has non-vanishing momentum. It is easy
to   imagine  three dispersing modes in the UHB with a single hole 
which might be relevant  to the RIXS physics, a positively
charged spin polaron related with the hole, a negatively charged
spin polaron related with the DO and a bound state of these two
entities (exciton). Our calculation will clarify the mechanism of
formation and propagation of such objects.

Since we are concerned with the the Hubbard model in the large-$U$
limit we choose a standard way of separating Hubbard bands by
means of a unitary transformation. A version of an effective
$t$-$J$ model obtained in this way, which acts in the UHB was derived
by Eskes  
{\it et al.} \cite{Eskesetal1994}. Physical processes contained in
that model are actually very similar to the processes
which are present in the standard $t$-$J$ model defined in the
LHB, but involvement of doubly occupied sites prohibits derivation
of a compact form for the Hamiltonian in the case of the UHB. The effective
model for the UHB promotes AF
correlations, like its counterpart for the LHB,  and therefore we
choose the N\'eel state $|N\rangle$ as a point of reference
by means of which we are going to present the Hamiltonian in a
comprehensive way. The virtue of this representation is that it
demonstrates the crucial duality between a hole and a DO. For the purpose
of this paper we traditionally neglect in the effective Hamiltonian
terms related with correlated hopping if 
they do not involve simultaneously a hole and a DO. Let
$\sigma(i)$ denote the direction of spin at the site $i$ in the
N\'eel state $|N\rangle$ and $\bar{\sigma}(i)\equiv - \sigma(i)$.
States spanned by the fermionic Hilbert space may be obtained
by acting on $|N\rangle$ with products of
$c^{\dag}_{\bar{\sigma}(i)}c_{\sigma(i)}\equiv
\tilde{s}^{\dag}_i$, $c_{\sigma(i)}\equiv \tilde{h}^{\dag}_i$ or
$c^{\dag}_{\bar{\sigma}(i)}\equiv \tilde{d}^{\dag}_i$. They invert
a spin (create a magnon), create a hole or a DO at the site $i$.
The representation in terms of products will be unique if
each index $i$, which refers a to site, appears in all products no
more than once. Since  the effective
Hamiltonian obtained by Eskes {\it et al.} corresponds to the
second-order  
perturbation expansion it acts maximally on three sites
which lie in row in the lattice and hence  affects only three
operators in the above mentioned products. For the same reason
only a single hole-creation operator and a single DO-creation
operator may be involved in the action of the effective
Hamiltonian. In (\ref{beg}-\ref{en}) the action of the $t$-$J$ model
in the UHB with a single hole is represented by means of states which are
coupled by this operator. Relations  (\ref{beg}-\ref{en})
actually demonstrate 
how the action of the effective Hamiltonian interchanges pairs
or trios of creation operators represented by $\hat{O}$ and
$\hat{O}^\prime$ between product 
representations like $\hat{A} \hat{O} \hat{B} |N\rangle $ and
$\hat{A} \hat{O}^\prime \hat{B} |N\rangle $ of these states, 
($\hat{A}$ and $\hat{B}$ denote products of
operators which  act on different sites than those involved in
the action of $\hat{O}$ and $\hat{O}^\prime$). Lack of an operator
for a given site $i$ in the product is formally represented 
by the identity operator $\tilde{\iota}_i$. In this case the site $i$
is occupied by a single spin which points in the direction
determined by the original N\'eel configuration.
The value of the matrix element of the Hamiltonian is given  in
(\ref{beg}-\ref{en}) after 
a colon. The energy of the N\'eel state is chosen as a point of reference
from which we count the diagonal contributions to the Hamiltonian.
A different description of relations  (\ref{beg}-\ref{en}) is
that they represent a sum over states  with factors given by
amplitudes presented after colons. That sum is created by  action of
the Hamiltonian on a given state represented by a product. 
Each component of the sum may be  obtained by exchanging operators on two
or three sites according to the rules represented by    (\ref{beg}-\ref{en}).
 Processes in (\ref{beg})
represent hopping, or in other words shifts of a hole or a DO
accompanied by 
creation or annihilation of a magnon. In (\ref{xy})
anti-parallel spins on a pair of nearest sites are simultaneously inverted.
(\ref{rnb}-\ref{rne}) are related with the rise in the contribution
to diagonal matrix elements of the Hamiltonian (which play the role of
potential energy), in comparison to the  
vanishing contributions from (\ref{neel}),
which appears if nearest sites are not occupied by parallel
spins. That rise is higher if such a pair is occupied by a hole
and a DO, which may be seen in the second term in (\ref{rne}). On
the other hand in such a case an additional process is possible
that may lower the energy, namely the exchange of the hole and the
DO which may be seen in (\ref{ex}). Processes described by
formulas (\ref{texb}-\ref{en}) represent exchange, rotations and
shifts of 
the hole-DO pair between nearest links which in some cases are
accompanied by a shift of a magnon.
\begin{eqnarray}
\tilde{h}^{\dag}_i \tilde{\iota}_{i+\delta} \leftrightarrow
\tilde{s}^{\dag}_i \tilde{h}^{\dag}_{i+\delta}:t; &\;\;\; \;&
\tilde{d}^{\dag}_i \tilde{\iota}_{i+\delta} \leftrightarrow
\tilde{s}^{\dag}_i \tilde{d}^{\dag}_{i+\delta}:t\label{beg}
\\ \tilde{\iota}_i
\tilde{\iota}_{i+\delta} &\leftrightarrow&  \tilde{s}^{\dag}_i
\tilde{s}^{\dag}_{i+\delta}:J/2\label{xy}
\\ \tilde{\iota}_i
\tilde{\iota}_{i+\delta} \leftrightarrow  \tilde{\iota}_i
\tilde{\iota}_{i+\delta}:0;&\;\; \;& \tilde{s}^{\dag}_i
\tilde{s}^{\dag}_{i+\delta} \leftrightarrow  \tilde{s}^{\dag}_i
\tilde{s}^{\dag}_{i+\delta}:0 \label{neel}\\ \tilde{\iota}_i
\tilde{s}^{\dag}_{i+\delta} \leftrightarrow \tilde{\iota}_i
\tilde{s}^{\dag}_{i+\delta}:J/2;&\;\; \; &\tilde{\iota}_i
\tilde{h}^{\dag}_{i+\delta} \leftrightarrow \tilde{\iota}_i
\tilde{h}^{\dag}_{i+\delta}:J/2 \label{rnb}\\  \tilde{s}^{\dag}_i
\tilde{h}^{\dag}_{i+\delta} \leftrightarrow  \tilde{s}^{\dag}_i
\tilde{h}^{\dag}_{i+\delta}:J/2;&\;\; \;& \tilde{\iota}_i
\tilde{d}^{\dag}_{i+\delta} \leftrightarrow \tilde{\iota}_i
\tilde{d}^{\dag}_{i+\delta}:J/2
\\  \tilde{s}^{\dag}_i
\tilde{d}^{\dag}_{i+\delta} \leftrightarrow  \tilde{s}^{\dag}_i
\tilde{d}^{\dag}_{i+\delta}:J/2;&\;\; \;& \tilde{h}^{\dag}_i
\tilde{d}^{\dag}_{i+\delta} \leftrightarrow \tilde{h}^{\dag}_i
\tilde{d}^{\dag}_{i+\delta}:J \label{rne}\\  \tilde{h}^{\dag}_i
\tilde{d}^{\dag}_{i+\delta} &\leftrightarrow&  \tilde{d}^{\dag}_i
\tilde{h}^{\dag}_{i+\delta}:-J/2 \label{ex}\\  \tilde{h}^{\dag}_i
\tilde{d}^{\dag}_{i+\delta} \tilde{\iota}_{i+\delta+\delta^\prime}
&\leftrightarrow& \tilde{\iota}_i \tilde{h}^{\dag}_{i+\delta}
\tilde{d}^{\dag}_{i+\delta+\delta^\prime}:J/4 \label{texb}
\\  \tilde{h}^{\dag}_i
\tilde{d}^{\dag}_{i+\delta}
\tilde{s}^{\dag}_{i+\delta+\delta^\prime} &\leftrightarrow&
\tilde{s}^{\dag}_i \tilde{h}^{\dag}_{i+\delta}
\tilde{d}^{\dag}_{i+\delta+\delta^\prime}:J/4\\ \tilde{h}^{\dag}_i
\tilde{d}^{\dag}_{i+\delta} \tilde{\iota}_{i+\delta+\delta^\prime}
&\leftrightarrow& \tilde{\iota}_i \tilde{d}^{\dag}_{i+\delta}
\tilde{h}^{\dag}_{i+\delta+\delta^\prime}:-J/4
\\  \tilde{h}^{\dag}_i
\tilde{d}^{\dag}_{i+\delta}
\tilde{s}^{\dag}_{i+\delta+\delta^\prime} &\leftrightarrow&
\tilde{s}^{\dag}_i \tilde{d}^{\dag}_{i+\delta}
\tilde{h}^{\dag}_{i+\delta+\delta^\prime}:-J/4\\
\tilde{d}^{\dag}_i \tilde{h}^{\dag}_{i+\delta}
\tilde{\iota}_{i+\delta+\delta^\prime} &\leftrightarrow& \tilde{\iota}_i
\tilde{h}^{\dag}_{i+\delta}
\tilde{d}^{\dag}_{i+\delta+\delta^\prime}:-J/4
\\  \tilde{d}^{\dag}_i
\tilde{h}^{\dag}_{i+\delta}
\tilde{s}^{\dag}_{i+\delta+\delta^\prime} &\leftrightarrow&
\tilde{s}^{\dag}_i \tilde{h}^{\dag}_{i+\delta}
\tilde{d}^{\dag}_{i+\delta+\delta^\prime}:-J/4 \label{en}
\end{eqnarray}
The duality between the hole and the DO may be clearly seen in
(\ref{beg}-\ref{en}). We may also notice the similarity with the
$t$-$J$ model acting in LHB. The main differences are that the
rise of the potential energy is only $J/2$ if a hole pair occupies
nearest sites and not $J$ like for a hole-DO pair and that in the LHB
only one kind of charged particles (holes) is present.
Nevertheless, one may expect that the possible mechanism of
binding of a hole and a DO in the UHB is similar to the mechanism
of hole pairing in the LHB.

\begin{figure}
 \unitlength1cm
\begin{picture}(8.0,4.2)
\epsfxsize=8.0cm
\put(0.0,-4.55){\epsfbox{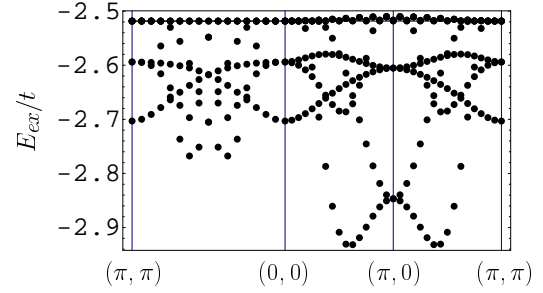}}
\end{picture}
\caption{Energy dispersion of eight lowest states in the UHB with a
single hole on a cluster of $32 \times 32$ sites for $J/t=0.4$.}
\label{disper}
\end{figure}

Much is already known about binding of holes in weakly doped
antiferromagnets \cite{WrobelEder98}. We will use these insights
to analyze formation of an exciton in the UHB. Fast motion, with
the rate $\sim t$,  mediated by  terms (\ref{beg}) of a hole or a
DO created at some site in the N\`eel state produces magnons which
lie on the track of the hole or the DO and contribute to the rise in
the potential energy (terms 
(\ref{rnb}-\ref{rne})). Strings consisting of  magnons are pinned to
the site where these objects have been initially created and
form  a potential well for a hole and a DO which confines their
motion. Much slower processes related to the process of inverting 
a pair of 
anti-parallel spins at nearest neighbor sites (process (\ref{xy})) 
at the begin of the string, which occur at the rate $\sim J$
shorten strings and give rise to coherent propagation of both 
objects. Since the separation of energy scales for hopping of a
hole or  a DO and annihilation of defects (magnons) is pronounced
it is plausible to introduce the notion of a spin polaron. We may
define in this context a hole or a DO spin-polaron as a solution of
the Schr\"odinger equation for a particle in the potential well,
the wave function of which is  as a combination of states that may be
reached by hopping of a hole (or a DO) created at an initial site $i$,
\begin{equation}
|\Psi^{H(DO)}_{ i }\rangle = \sum_{{\cal P}_i} \alpha_{{\cal P}_i}
|{\cal P}^{H(DO)}_i\rangle. \label{wvfnsp}
\end{equation}
${\cal P}_i$ parameterizes the geometry of a path along which the
hole or a DO has been moving and $|{\cal P}^{H(DO)}_i\rangle$ is a
state which has been created in this way. To be precise, at this
stage of our considerations we analyze an "unperturbed" Hamiltonian
which 
consists of terms 
symbolized  by relations (\ref{beg}) and (\ref{neel}-\ref{rne})
and represents fast oscillations of a hole or a DO in the vicinity of
the initial site. Slower processes are neglected in this phase of
analysis. ``Orbital'' states of polarons created at all possible
sites exhaust, in principle, the relevant portion of the Hilbert
space. In practice, the calculation may be confined to some low
excited states or even to the polaron ground state. The concept of
spin polarons may be extended to the description of interaction
between a hole and a DO.  If we consider the hole and the DO which
have been initially created at distant sites we just apply the
generalization of a single particle problem by assuming that the
wave function in this case is a product of wave functions for a
hole and a DO. For a pair of particles created an nearest-neighbor
sites $\langle i,j \rangle$ we define a hole-DO spin-bipolaron
$|\Psi^{HDO}_{\langle i,j \rangle }\rangle$ as a sum over states
which may be reached by independent hopping of a hole and a DO
created at the sites $i$ and $j$,
\begin{equation}
|\Psi^{HDO}_{\langle i,j \rangle }\rangle = \sum_{{\cal P}_i,{\cal
P}_j} \alpha_{{\cal P}_i,{\cal P}_j} |{\cal P}^H_i,{\cal
P}^{DO}_j\rangle. \label{wvfnsbp}
\end{equation}
The meaning of symbols is the same as in the case of single
particles.  At this stage of considerations we prohibit, by
definition, the hole and the DO to follow along the trace left by
the accompanying particle and solve the Schr\"odinger equation in
this restricted Hilbert space for an "unperturbed" Hamiltonian,
which takes into account only hopping of the hole or the DO and
the  diagonal "potential" contribution to the energy from the
magnetic $J$-term in the $t$-$J$ model. By means of that 
restriction we achieve that spin bipolarons are
localized and we may proceed as in the case of a single hole. We
make a further approximation and neglect some path details including
the possibility of path crossing. It is self-evident that the
coefficients $\alpha_{{\cal P}_i}$ and $\alpha_{{\cal P}_i,{\cal
P}_j}$ for  polarons in the ground state will  depend only on the
lengths $\mu$, $\nu$ of paths ${\cal P}_i$ and ${\cal P}_j$, and thus
$\alpha_{{\cal
  P}_i}=\alpha_{\mu}$ and
$\alpha_{{\cal
  P}_i,{\cal P}_j}=\alpha_{\mu,\nu}$.
By calculating the matrix elements of the full Hamiltonian in the
polaron basis we derive an effective Hamiltonian expressed in the
polaron language. Some processes, like oscillations of the hole
or the DO in the vicinity of a polaron center, are already incorporated
into the eigenenergy of polaron states. The rest   which we
discuss now   either renormalizes the eigenenergies of
polarons or gives rise to off-diagonal matrix elements in the
polaron Hamiltonian. The latter circumstance occurs  in the case
of shortening of strings attached to a single particle. This process
governs 
propagation of a single hole or a single DO in an
antiferromagnetic spin background. Formation of bipolarons, and,
in particular, strings that connect a hole and a DO, is an
effective way of lowering the energy. A compromise between two
opposite tendencies to minimize the kinetic energy of holes and
DOs and to reduce the disturbance of the antiferromagnetic
background is reached in this way. By shrinking at one end and
expanding at the opposite end, a string may move, while keeping a
moderate length. The physics of interaction between a hole and a DO is
richer than 
just for two holes, because it incorporates processes described by
relations (\ref{ex}-\ref{en}), but the method of deriving the 
Hamiltonian in the polaron language is the same as for one kind of
particles\cite{WrobelEder98}. Instead of dwelling on details we
present now an explicit form of the effective Hamiltonian $H_{eff}$.
$h^{\dag}_i$ and $d^{\dag}_i$ are fermionic operators which  create
a hole and the DO spin-polaron at the site $i$. They obey the relation
$h^{\dag}_id^{\dag}_i=0$ because none of  sites may be
simultaneously occupied by a hole and a DO. The bipolaron
$|\Psi^{HDO}_{\langle i,j \rangle }\rangle$ is created by the pair
$h^{\dag}_id^{\dag}_j$, where $i$ and $j$ are nearest-neighbor sites.
\begin{eqnarray}
&H_{eff}&= E_1  \sum_i (h^{\dag}_i h_i+ d^{\dag}_i d_i)+2
\chi_1 \sum_{i,\delta,\delta^\prime:\delta^\prime\neq-\delta}
h^{\dag}_{i+\delta+\delta^\prime} h_i\nonumber\\&+&2 \chi_1
\sum_{i,\delta,\delta^\prime:\delta^\prime\neq-\delta}
d^{\dag}_{i+\delta+\delta^\prime} d_i +(E_2-2 E_1) \nonumber\\
&\times&\sum_{i,\delta}d^{\dag}_{i+\delta} d_{i+\delta} h^{\dag}_i
h_i+(E_2 \omega_1+\tau_1+\iota/2)
\nonumber\\&\times&\sum_{i,\delta,\delta^\prime:\delta^\prime\neq\delta}
(d^{\dag}_{i} d_{i+\delta^\prime} h^{\dag}_{i+\delta} h_i+H.c.)+(
E_2 \omega_2+\tau_2)\nonumber\\ &\times&
\sum_{i,\delta,\delta^\prime,\delta^{\prime\prime}:
\delta^\prime\neq-\delta,\delta^{\prime\prime}\neq-\delta^\prime}
(d^{\dag}_{i+\delta+\delta^\prime} d_{i}
h^{\dag}_{i+\delta+\delta^\prime+\delta^{\prime\prime}}
h_{i+\delta}+H.c.)\nonumber\\ &+&\iota
\sum_{i,\delta}d^{\dag}_{i+\delta} d_{i+\delta}
h^{\dag}_{i+\delta} h_i +\iota/2\sum_{i,\delta,\delta^\prime:
\delta^\prime\neq\delta}d^{\dag}_{i} d_{i} h^{\dag}_{i+\delta}
h_{i+\delta^\prime}\nonumber\\&+&\iota/2\sum_{i,\delta,\delta^\prime:
\delta^\prime\neq\delta}d^{\dag}_{i+\delta} d_{i+\delta^\prime}
h^{\dag}_{i}
h_{i}\nonumber\\&-&2\chi_1\sum_{i,\delta,\delta^\prime:
\delta^\prime\neq-\delta}d^{\dag}_{i+\delta} d_{i+\delta}
h^{\dag}_{i+\delta+\delta^\prime} h_{i}\nonumber\\
&-&2\chi_1\sum_{i,\delta,\delta^\prime:
\delta^\prime\neq-\delta}d^{\dag}_{i+\delta+\delta^\prime} d_{i}
h^{\dag}_{i+\delta} h_{i+\delta}\nonumber\\&+&\chi_2
\sum_{i,\delta,\delta^\prime,\delta^{\prime\prime}:
\delta^\prime\neq\delta,\delta^{\prime\prime}\neq-\delta^\prime}
(d^{\dag}_{i+\delta+\delta^{\prime\prime}} d_{i+\delta}
h^{\dag}_{i+\delta^\prime} h_{i}+H.c.).\nonumber \\ \label{ham}
\end{eqnarray}
$E_1$ and $E_2$ denote the eigenenergies of the spin polaron and
the bipolaron which represent the energy of a particle or
two particles pinned to starting-point sites by strings, while the rest
of parameters is either related with overlap between polaron
states at different sites  or processes
that couple polaron states:
\begin{eqnarray}
\omega_1&=&\sum_{\mu=0,\nu=1}(z-1)^{\mu+\nu-1} \alpha_{\nu,\mu}
\alpha_{\nu-1,\mu+1},\\
\omega_2&=&\sum_{\mu=0,\nu=2}(z-1)^{\mu+\nu-2} \alpha_{\nu,\mu}
\alpha_{\nu-2,\mu+2},\\ \tau_1&=&t\sum_{\mu=1}(z-1)^{\mu-1}
\alpha_{\mu,0} \alpha_{\mu-1,0},\\
\tau_2&=&t\sum_{\mu=1}(z-1)^{\mu-2} \alpha_{\mu,0}
\alpha_{\mu-2,0},\\ \chi_1&=&J/2\sum_{\mu=2}(z-1)^{\mu-2}
\alpha_{\mu} \alpha_{\mu-2},\\
\chi_2&=&J/2\sum_{\mu=1,\nu=1}(z-1)^{\mu+\nu-2} \alpha_{\nu,\nu}
\alpha_{\mu-1}\alpha_{\nu-1},\\ \iota&=&J/2 (\alpha_{0,0})^2.
\end{eqnarray}
As we have mentioned before polaron states are not necessarily
orthogonal and in order to find eigenenergies of the system
one has to diagonalize the operator $O^{-1}_{eff} H_{eff}$, where
the operator $O_{eff}$ represents overlap between polaron states,
\begin{eqnarray}
&O_{eff}&=1+ \omega_1
\sum_{i,\delta,\delta^\prime:\delta^\prime\neq\delta}
(d^{\dag}_{i} d_{i+\delta^\prime} h^{\dag}_{i+\delta} h_i+H.c.)+
\omega_2\nonumber\\ &\times&
\sum_{i,\delta,\delta^\prime,\delta^{\prime\prime}:
\delta^\prime\neq-\delta,\delta^{\prime\prime}\neq-\delta^\prime}
(d^{\dag}_{i+\delta+\delta^\prime} d_{i}
h^{\dag}_{i+\delta+\delta^\prime+\delta^{\prime\prime}}
h_{i+\delta}+H.c.). \nonumber\\ \label{overl}
\end{eqnarray}
The Hamiltonian (\ref{ham}) and the overlap operator (\ref{overl})
for a hole and a DO are very similar to operators that represent
interaction of holes in the LHB. Thus the underlying
mechanisms of exciton formation, if it occurs, and pairing of
holes  should be very similar. Fig.\ref{disper}  depicts the
energy dispersion of eight states with lowest energies obtained by
solving the eigenvalue problem for the operators (\ref{ham}) and
(\ref{overl}) in the case of a single hole and a single DO. 
The line in the upper part of Fig.\ref{disper}  represents the
lower boundary for the energy of scattering states. There exist
three bound excitonic states at the wave vector ${\bf k}=(0,0)$.
The ground state has $d$-wave symmetry. The pair of bound states with 
higher energy has $p$-wave symmetry, which guarantees that they
may take part in dipole transitions and contribute to Raman
spectra.
\noindent 
For
other values of the wave vector up to five exciton states at
different energies may exist. The typical distance between the
hole spin-polaron and the DO spin-polaron in the exciton state
is  one lattice spacing which indicates that they are tightly bound
and 
constitute  a bipolaron. Among dispersion curves which form in
Fig.\ref{disper} one may identify  bands which shape agree with the form
of energy-dispersion curves for the particle-hole excitation measured by Hasan
{\it et 
al.} in the RIXS
experiment, but the bandwidth in the theoretical approach for a
reasonable hopping 
parameter 
$t=0.35~eV$ seems to be much to small, which should be attributed
to the fact that we have neglected the possibility of direct hopping to
further neighbors. Additional analysis is also needed to understand
the dependence of spectral intensity  on the position of the wave
vector in the Brillouin zone.

One of the authors (P.W.) acknowledges support
by the Polish Science Committee (KBN) under contract No. 5 P03B
058 20.

\end{document}